\PassOptionsToPackage{hyphens}{url}
\documentclass[aps,prl,reprint,longbibliography,floatfix]{revtex4-2}
\usepackage{amsmath,amssymb,graphicx}

\begin{document}

\title{Solution of the Qubit Marginal Problem from Fermionic Representability}

\author{David A. Mazziotti}
\email{damazz@uchicago.edu}
\affiliation{Department of Chemistry and The James Franck Institute, The University of Chicago, Chicago, Illinois 60637, USA}

\date{Submitted November 13, 2026}

\begin{abstract}
Whether local reduced states are compatible with a global quantum state---the
quantum marginal problem---is central to quantum information and simulation.
Here we present a solution to the ensemble marginal problem for qubits based on the
representability theory of fermions. Encoding each qubit as one fermion in two orbitals,
we prove that the representable qubit marginals form an exposed face of the set of
$N$-representable fermionic reduced density matrices, selected by the vanishing of every
site-occupation variance. Complete hierarchies of fermionic representability
conditions therefore transfer exactly to qubits. Applications to frustrated Heisenberg
and transverse-field Ising rings show that the $(2,3)$-positivity conditions are
exact at the Majumdar--Ghosh point and accurate to about 0.5\% at criticality. Because
the conditions hold for any quantum state, they also provide a framework for
mitigating measurement errors on qubit-based quantum devices.
\end{abstract}

\maketitle

\noindent\textit{Introduction.}---Whether a collection of local reduced density matrices
can arise from a single global quantum state is the quantum marginal
problem~\cite{Klyachko06,Schilling14}. For qubits it underlies the
characterization of multipartite entanglement~\cite{Higuchi03}, the
certification of data from quantum devices, and every quantum simulation in
which the measured quantities are few-body marginals rather than the full
state. In general the problem is QMA-complete~\cite{Liu06,LCV07}, so efficient
characterizations must take the form of hierarchies of necessary conditions,
and the central questions are which conditions to impose and whether the
hierarchy converges. For pure states, hierarchies have been formulated that
impose purity on the global state, at a cost that grows exponentially with the
number of qubits~\cite{Yu21}.

For fermions, the same question is the $N$-representability problem: the
conditions under which a two-particle reduced density matrix (2-RDM) derives
from an $N$-particle
state~\cite{Coleman63,GarrodPercus64,Kummer67,Erdahl78}. A complete hierarchy
of $N$-representability conditions has been
constructed~\cite{Mazziotti12,Mazziotti12PRA,Mazziotti23}, in which conditions on
higher-order RDMs are converted into conditions on the 2-RDM alone by
cancelling all higher-body contributions, and it has recently been extended to
systems without particle-number conservation~\cite{Mazziotti26}. These
conditions enable direct variational calculations of the 2-RDM with rigorous
lower bounds on the energy. Such calculations have been realized through
approximate representability conditions with significant applications to
strongly correlated molecules and
materials~\cite{MazziottiErdahl01,Nakata01,Mazziotti02,Mazziotti04,Zhao04,Cances06,Shenvi10,Poelmans15,Mazziotti16,Alcoba18,Mazziotti20,Knight22,Knight25,Schouten25},
including applications from conduction in glassy
polymers~\cite{Xie22} to exciton condensation in molecular
materials~\cite{Safaei18,Torres24} (see also
Refs.~\cite{Kais07,Schouten26,Piris21,BenavidesRiveros20} for connections with
entanglement and one-particle RDM functional theories).

In this Letter we present a solution of the quantum marginal problem for qubits,
which we call the ensemble qubit representability problem, from the theory of
fermionic $N$-representability. Mapping each qubit onto one fermion in two
orbitals~\cite{Abrikosov65,Schwerdtfeger09,LCV07,Chakraborty15,Smart19}, we
prove that the representable $p$-qubit marginals form an exposed face of the
set of $N$-representable $p$-particle reduced density matrices ($p$-RDMs),
selected by the vanishing of every site-occupation variance, and hence complete
fermionic hierarchies transfer exactly to qubits. In a Majorana basis, Pauli weight
equals fermionic body order, such that in the generation of the representability
conditions the cancellation of higher-body terms corresponds to the
cancellation of high-weight Pauli strings. Because every qubit operator
conserves each site occupation, the resulting semidefinite programs separate into
polynomially many small blocks. We apply the
$(2,2)$- and $(2,3)$-positivity conditions to frustrated Heisenberg and
transverse-field Ising rings and discuss their application beyond ground states,
including to error mitigation in quantum simulation.

\noindent\textit{Qubit representability as a face of $N$-representability.}---We
represent each of the $N$ qubits by one fermion in two orbitals. With fermionic
operators on site $s$, the Pauli operators become one-particle operators confined to
that site,
\begin{equation}
  \hat\sigma^{\lambda}_{s}=\sum_{\mu\nu}\hat a^\dagger_{s\mu}\,(\sigma^{\lambda})_{\mu\nu}\,\hat a^{}_{s\nu},
  \qquad \lambda\in\{x,y,z\},
  \label{eq:pauli}
\end{equation}
where ${\hat a}^\dagger_{s\mu}$ and ${\hat a}^{}_{s\nu}$ are creation and annihilation
operators.  These Pauli operators reproduce the qubit algebra exactly on the subspace
$\mathcal{Q}$ of states in which every site holds exactly one fermion. That
subspace is isomorphic to the Hilbert space of $N$ qubits, and operators on
different sites commute because each is even in the fermionic operators. Every
Pauli string acting on $p$ sites is a product of $p$ such one-particle
operators, and hence a $p$-body operator. Consequently, every $p$-qubit
marginal is contained in the $p$-RDM of $N$ fermions in $r=2N$ orbitals, and
the qubit representability problem can be posed as a fermionic representability
problem.

The subspace $\mathcal{Q}$ is selected by the site-occupation variance operators
\begin{equation}
  \hat Q_s=(\hat n_s-1)^2,\qquad
  \hat n_s=\hat a^\dagger_{s\alpha}\hat a^{}_{s\alpha}+\hat a^\dagger_{s\beta}\hat a^{}_{s\beta}.
  \label{eq:variance}
\end{equation}
Because $\hat n_s$ has eigenvalues 0, 1, and 2, each $\hat Q_s$ is a positive
semidefinite two-body operator that annihilates exactly the states in which
site $s$ is singly occupied, and $\mathcal{Q}$ is the null space of
$\hat Q=\sum_s\hat Q_s$. For $p\ge2$ we define
\begin{equation}
  F^p_N=\big\{\,{}^pD\in P^p_N:\ \mathrm{Tr}\big(\hat Q\,{}^pD\big)=0\,\big\},
  \label{eq:face}
\end{equation}
where $P^p_N$ is the convex set of $N$-representable
$p$-RDMs~\cite{Coleman63,Kummer67}.

\textit{Theorem.}---\textit{For $2\le p\le N$, $F^p_N$ is an exposed face of
$P^p_N$ whose elements are exactly the $p$-RDMs of $N$-qubit states. Because
each such $p$-RDM is linearly equivalent to the $p$-qubit marginals of its
state, the set of representable $p$-qubit marginals is the face $F^p_N$ of the
set of $N$-representable $p$-RDMs.}

\textit{Proof.}---(i) Because $\hat Q\succeq0$ and $\mathrm{Tr}(\hat Q\,{}^pD)\ge0$
on $P^p_N$, the operator $\hat Q$ defines the supporting hyperplane
$\mathrm{Tr}(\hat Q\,{}^pD)=0$ of $P^p_N$, and $F^p_N$ is an exposed
face~\cite{Rockafellar70}. (ii) Any $N$-particle density matrix ${}^ND$ whose
$p$-RDM lies in $F^p_N$ has $\mathrm{Tr}(\hat Q\,{}^ND)=0$, or
$\hat Q\,{}^ND=0$, and hence is an $N$-qubit state. Conversely, every
$N$-qubit state is an $N$-fermion state on $\mathcal{Q}$ whose $p$-RDM lies in
$F^p_N$. (iii) On $\mathcal{Q}$, a $p$-RDM element $\langle\hat O\rangle$, where
$\hat O$ is a product of $p$ creation and $p$ annihilation operators, vanishes
if $\hat O$ changes any site occupation. Otherwise, because $\hat O$ commutes
with the bilinears of Eq.~\eqref{eq:pauli} on every site not involved in
$\hat O$, it acts on at most $p$ qubits as a combination of Pauli strings of
weight at most $p$. Conversely, a Pauli string of weight $w\le p$ is a
$w$-body operator, whose expectation value follows from ${}^pD$. Thus ${}^pD$
and the $p$-qubit marginals determine each other linearly.\hfill$\square$

By the bipolar theorem~\cite{Kummer67, Rockafellar70}, the face $F^2_N$ is determined by
its polar cone, the two-body operators with nonnegative expectation values on
$F^2_N$,
\begin{equation}
  (F^2_N)^*=\big\{\,{}^2\hat B:\ \hat P\,{}^2\hat B\,\hat P\succeq0\,\big\},
  \label{eq:polar}
\end{equation}
where $\hat P=\prod_s\hat n_s(2-\hat n_s)$ projects onto $\mathcal{Q}$.
Building on this nonconstructive characterization, the constructive solution of
$N$-representability~\cite{Mazziotti12,Mazziotti12PRA,Mazziotti23,Mazziotti26}
yields a complete hierarchy of qubit representability conditions by generating
this cone from within: the $(2,p)$-positivity conditions are obtained from
conic combinations of squares of degree-$p$ polynomials in the fermionic
operators in which all higher-than-two-body terms cancel on $\mathcal{Q}$, and
each such combination lies in $(F^2_N)^*$. At each level the primal and polar
descriptions are equivalent, with no duality gap~\cite{Erdahl07,Slater50}:
because the products of the generators span all two-body operators after
compression, $\hat H-E$ has a representation with a positive definite
coefficient matrix for sufficiently negative $E$, so Slater's condition holds.

\noindent\textit{Pauli strings as Majorana operators.}---The cancellation of
higher-body terms takes a simple form in the Majorana-like operators
$\hat p_k=\hat a^{}_k+\hat a^\dagger_k$ and $\hat m_k=\hat a^{}_k-\hat a^\dagger_k$ of
Ref.~\cite{Mazziotti26}. We write them as the Hermitian operators
$\hat\gamma_{2k-1}=\hat p_k$ and $\hat\gamma_{2k}=-i\hat m_k$, which satisfy
$\{\hat\gamma_\mu,\hat\gamma_\nu\}=2\delta_{\mu\nu}$, so that each site $s$
carries four Majorana operators. Products of Majorana operators are again
Majorana monomials up to a phase. We multiply operators in the full Fock space
and compare them on $\mathcal{Q}$ only through their compressions
$\hat P\hat M\hat P$. Compression is linear but not multiplicative, so the
operators that vanish on $\mathcal{Q}$, those with $\hat P\hat A\hat P=0$, may
be dropped from a sum but not from a factor of a product. On $\mathcal{Q}$, the
product of the four operators on a site equals the identity, because the site
parity is fixed by single occupancy. Hence a product of two Majorana operators
on a site has, up to a phase, the same compression as the product of the other
two, and every Pauli operator in Eq.~\eqref{eq:pauli} has the same compression
as a single Majorana bilinear; for example,
$\hat\sigma^z_s=i\hat\gamma_{s1}\hat\gamma_{s2}$ on $\mathcal{Q}$, where
$\hat\gamma_{s1},\dots,\hat\gamma_{s4}$ denote the Majorana operators of site
$s$. A Pauli string acting on $w$ sites therefore has the same compression as a
Majorana monomial of degree $2w$, that is, a $w$-body operator.  Conversely,
the compression of every Majorana monomial is either zero, when the monomial
contains an odd number of Majorana operators on some site and hence changes
that site's occupation, or a Pauli string up to a phase. Compression thus
identifies the Majorana monomials with the Pauli strings, and the body order of
an operator becomes the weight of the corresponding Pauli string.

This dictionary converts the fermionic cancellation into a cancellation of
Pauli strings. In the $(2,p)$-positivity conditions, conic combinations of
squares of degree-$p$ polynomials in the Majorana operators are written in the
Gram form $\sum_{IJ}X_{IJ}\,\hat O_I^{}\hat O_J^\dagger$, where
the $\hat O_I^{}$ are Majorana monomials of degree at most $p$ and $X$ is a
positive semidefinite matrix. Because the Majorana operators square to the
identity and anticommute, the product $\hat O_I^{}\hat O_J^\dagger$ is a single
monomial up to a phase, and its compression is either zero or a single Pauli
string. Each element of $X$ therefore contributes to exactly one Pauli
string, and the requirement that the combination be a two-body operator on
$\mathcal{Q}$ becomes the requirement that all Pauli strings of weight greater
than two cancel. Imposing the cancellation modulo the operators that vanish on
$\mathcal{Q}$ enforces both the vanishing site variances and the contracted
Schr\"odinger equations of the site-occupation operators,
$\mathrm{Tr}\big[{}^p\hat\Gamma\,(\hat n_s-1)\,{}^ND\big]=0$ for all $p$-particle
operators ${}^p\hat\Gamma$, so that no further conditions on the site
occupations are required.

For a two-local qubit Hamiltonian $\hat H$, the lower bound on the ground-state
energy from the $(2,p)$-positivity conditions is obtained from the
semidefinite program~\cite{VandenbergheBoyd96}
\begin{align}
  &\max_{E,\;{}^2\hat B,\;X\succeq 0}\ E \label{eq:sdp}\\
  &\text{such that}\quad \hat H-E-{}^2\hat B\equiv 0, \label{eq:sdp-two}\\
  &\phantom{\text{such that}}\quad
   {}^2\hat B-\sum_{IJ}X_{IJ}\,\hat O_I^{}\hat O_J^\dagger\equiv 0,
  \label{eq:sdp-constraint}
\end{align}
where $\equiv$ denotes equality after compression to $\mathcal{Q}$ and ${}^2\hat B$ is a
two-body operator, meaning that it is restricted to Pauli strings of weight at
most two. The first equation
identifies $\hat H-E$ with the two-body operator ${}^2\hat B$, and its
Lagrange multiplier is the 2-RDM, here the 2-qubit marginals~\cite{Mazziotti20,
Mazziotti23,Mazziotti26}. The second equation places ${}^2\hat B$ in the polar
cone generated from conic combinations of squares: expanding it in Pauli strings
yields one linear equation per string, in which the strings of weight at most two match
${}^2\hat B$ and those of higher weight cancel. The Lagrange multipliers
of the second equation form the corresponding $p$-positive RDM: stationarity
with respect to ${}^2\hat B$ makes those of the strings of weight at most two
equal to the 2-RDM, while those of the cancellation equations, which need not
be computed, are the remaining, higher-weight elements. Hamiltonians with
$k$-body interactions are treated in the same way with the cancellation of
Pauli strings of weight greater than $k$.

\noindent\textit{Site-occupation symmetry.}---Each qubit is a single fermion that
occupies one of the two orbitals on its site, and by Eq.~\eqref{eq:pauli} every
qubit operator commutes with every site occupation $\hat n_s$. The qubit problem
therefore carries a symmetry that the electronic-structure problem lacks: in
place of the single conserved particle number, there are $N$ conserved site
occupations. The squares in the Gram form,
however, are built from fermionic operators such as $\hat a_{s\alpha}$ or
$\hat a^\dagger_{s\beta}$, which acting on a qubit state temporarily empty a
site or doubly occupy it. We classify each generator by the change $\Delta_s$
that it produces in the occupation of every site,
\begin{equation}
  [\hat n_s,\hat O_I]=\Delta_s(I)\,\hat O_I .
  \label{eq:charge}
\end{equation}
Averaging Eqs.~\eqref{eq:sdp-two}--\eqref{eq:sdp-constraint} over the phase rotations
$\exp(i\sum_s\theta_s\hat n_s)$ leaves $\hat H$ invariant and maps squares of
degree-$p$ polynomials into squares of the same degree, so the Gram matrix $X$
can be chosen invariant under these rotations. Its elements then vanish between
generators with different patterns $\Delta=(\Delta_1,\dots,\Delta_N)$, and $X$
becomes block diagonal, with one block for each pattern. This structure, noted for
the transverse-field Ising model in Ref.~\cite{Schwerdtfeger09}, here holds at
every level and for any qubit Hamiltonian. In approaches
formulated directly with Pauli operators, by contrast, every operator conserves
the site occupations, so this decomposition does not arise; it appears here
because the fermionic generators temporarily leave $\mathcal{Q}$. Only the changes
$\Delta_s\in\{-1,0,+1\}$ survive, because a singly occupied site can neither
lose nor gain two fermions. For $p=3$ there are $2N$ blocks of dimension
$6N-4$, in which one site changes occupation, and $8\binom{N}{3}$ blocks of
dimension 8, in which three sites do; for $N=12$, a Gram matrix of dimension
15\,712 thus separates into 24 blocks of dimension 68 and 1\,760 blocks of
dimension 8.

\noindent\textit{Partial conditions.}---A subset of the $(2,3)$-positivity conditions
requires no cancellation equations at all. For any polynomial $\hat O$ of odd
degree at most three, the anticommutator
$\hat O\hat O^\dagger+\hat O^\dagger\hat O$ contains no three-body terms, because
the three-body parts of the particle and hole squares cancel identically. These
are the generalized T1 and T2
conditions~\cite{Erdahl78,Zhao04,Braams07,Mazziotti12PRA,Mazziotti26}. Together with the
$(2,2)$-positivity conditions, they define the partial $(2,3)$-positivity
conditions, which involve only the Pauli strings of weight at most two. The
full $(2,3)$-positivity conditions, which also include the squares whose
three-body terms cancel only in combination, contain the partial conditions and
are in general strictly stronger, as found for fermions in
Ref.~\cite{Mazziotti26}.

\noindent\textit{Results.}---We apply the $(2,2)$-, partial $(2,3)$-, and full
$(2,3)$-positivity conditions to two periodic rings of $N=12$ qubits: the
frustrated Heisenberg ring with nearest- and next-nearest-neighbor couplings,
\begin{equation}
  \hat H=J_1\sum_{i}\hat{\mathbf S}_i\cdot\hat{\mathbf S}_{i+1}
        +J_2\sum_{i}\hat{\mathbf S}_i\cdot\hat{\mathbf S}_{i+2},
  \label{eq:heis}
\end{equation}
with $\hat{\mathbf S}_i=\hat{\boldsymbol\sigma}_i/2$ and $J_2/J_1$ varied between 0 and 1,
and the transverse-field Ising ring,
\begin{equation}
  \hat H=-J\sum_{i}\hat\sigma^z_i\hat\sigma^z_{i+1}-g\sum_{i}\hat\sigma^x_i,
  \label{eq:tfim}
\end{equation}
with $g/J$ varied between 0 and 2. The semidefinite programs of
Eqs.~\eqref{eq:sdp}--\eqref{eq:sdp-constraint}, in the block-diagonal form of the
site-occupation symmetry, are solved with the boundary-point
algorithm of Ref.~\cite{Mazziotti11}, with the solution at each value of the coupling used as
the starting point for the next. The ring size is chosen to enable direct
comparison with exact diagonalization and with the earlier variational 2-RDM
study of the transverse-field Ising ring~\cite{Schwerdtfeger09}; the cost of the
conditions, however, grows only polynomially with $N$ because of the
site-occupation symmetry. This scaling holds at each fixed level; convergence
may require levels that grow with $N$, as expected from
QMA-completeness~\cite{Liu06,LCV07}.

\begin{figure}[t]
  \centering
  \includegraphics[width=\columnwidth]{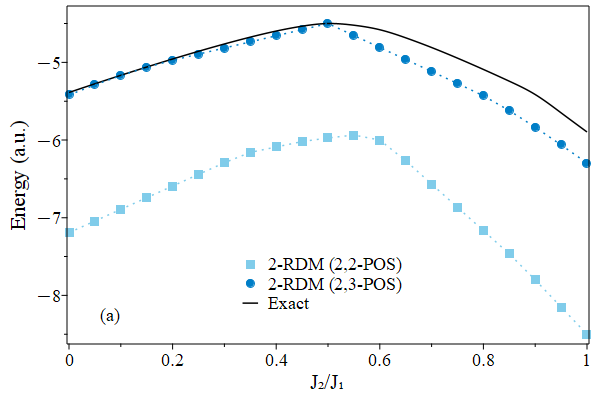}\\
  \includegraphics[width=\columnwidth]{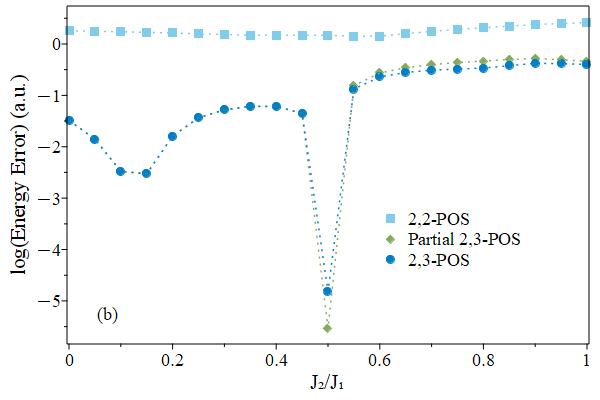}
  \caption{(a) Ground-state energy and (b) base-10 logarithm of the energy
  error $|E-E_{\rm exact}|$, in units of $J_1$, for the frustrated Heisenberg ring of
  12 qubits as a function of $J_2/J_1$, from the $(2,2)$-, partial $(2,3)$-, and
  full $(2,3)$-positivity conditions. The full conditions are exact at the
  Majumdar--Ghosh point $J_2/J_1=1/2$, where the error is limited by the
  convergence of the semidefinite program.}
  \label{fig:heis}
\end{figure}

Figure~\ref{fig:heis} shows the ground-state energy of the Heisenberg ring and
its error relative to exact diagonalization. The $(2,2)$-positivity conditions
underestimate the energy by $1.4$--$2.6\,J_1$, about a third of the exact
energy, across the entire range of couplings. The full $(2,3)$-positivity
conditions reduce the error to at most $0.06\,J_1$ for $J_2/J_1\le1/2$, with
errors of $3\times10^{-3}J_1$ near $J_2/J_1\approx0.1$, and they are
exact at the Majumdar--Ghosh point $J_2/J_1=1/2$~\cite{MajumdarGhosh69}. The
exactness has a transparent origin: at this point the Hamiltonian is a sum of
three-site terms,
$\hat H=\tfrac{J_1}{4}\sum_i(\hat{\mathbf S}_i+\hat{\mathbf S}_{i+1}+\hat{\mathbf S}_{i+2})^2
-\tfrac{9}{16}J_1 N$, each of which is minimized by the dimerized ground state,
and the positivity of the 3-qubit marginals contained in the $(2,3)$-positivity
conditions is sufficient to bound each term. On either side, the bound is linear in $J_2$ over
finite intervals, $0.25\lesssim J_2/J_1\le1/2$ and $1/2\le J_2/J_1\lesssim0.75$,
in each of which a single 2-RDM is optimal. In the
spiral regime $J_2/J_1>1/2$ the error of the full $(2,3)$ conditions increases to
$0.13$--$0.42\,J_1$, indicating that the incommensurate correlations of this regime require
higher levels of the hierarchy. The partial $(2,3)$-positivity conditions agree
with the full conditions to within $3\times10^{-5}J_1$ for $J_2/J_1\le1/2$ but
are less accurate in the spiral regime.

\begin{figure}[t]
  \centering
  \includegraphics[width=\columnwidth]{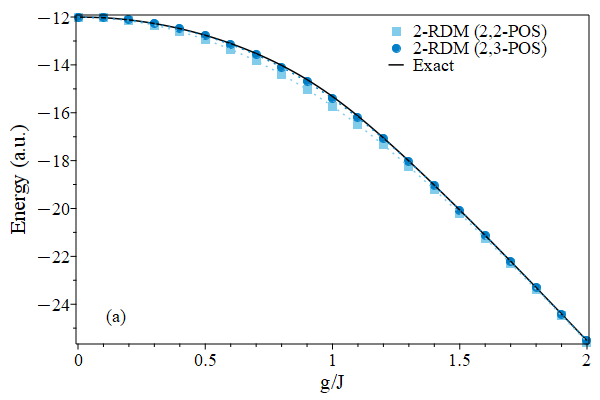}\\
  \includegraphics[width=\columnwidth]{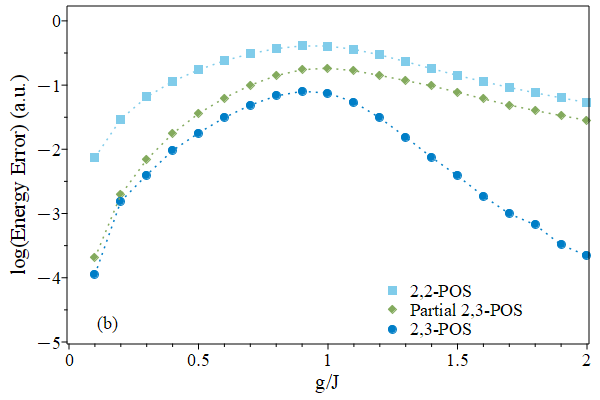}
  \caption{(a) Ground-state energy and (b) base-10 logarithm of the energy
  error $|E-E_{\rm exact}|$, in units of $J$, for the transverse-field Ising ring
  of 12 qubits as a function of $g/J$, from the $(2,2)$-, partial $(2,3)$-, and
  full $(2,3)$-positivity conditions. The point $g=0$, where all levels are
  exact, is omitted from (b).}
  \label{fig:tfim}
\end{figure}

Figure~\ref{fig:tfim} shows the corresponding results for the transverse-field
Ising ring. All three levels are exact at $g=0$, where the ground state is
classical, and their errors are largest near the critical field $g/J=1$. At the
critical point, where the exact energy is $-15.32\,J$, the $(2,2)$-, partial
$(2,3)$-, and full $(2,3)$-positivity conditions give energy errors of
$0.40\,J$, $0.18\,J$, and $0.075\,J$, respectively. The $(2,2)$ energies agree
with the 2-positivity energies of Ref.~\cite{Schwerdtfeger09} for the same
ring. The partial 3-positivity conditions employed there impose the E3 and F3
conditions on the 3-RDM, which are stronger than the T2 condition, and hence
give smaller errors than the partial $(2,3)$ conditions; nevertheless, the full
$(2,3)$ conditions reduce their error by a factor of two at the critical point,
from $0.16\,J$ to $0.075\,J$, and from $0.028\,J$ to $0.018\,J$ at
$g/J=0.5$. The distinction between the full and
partial conditions is most pronounced in the paramagnetic phase: at $g/J=2$ the
full $(2,3)$ conditions give an error of $2\times10^{-4}J$, two orders of
magnitude smaller than the error of $3\times10^{-2}J$ from the partial
conditions. As in the fermionic systems of Ref.~\cite{Mazziotti26}, the complete
intersection of the three-positive cone with the two-body space thus generates
qubit representability conditions beyond the T1 and T2 analogs.

\noindent\textit{Discussion and conclusions.}---We have shown that the qubit
representability problem is a face of the fermionic $N$-representability
problem, so that the complete and constructive hierarchy of
$(2,p)$-positivity conditions applies to qubits without modification. In a
Majorana basis the cancellation of higher-body terms becomes the cancellation
of high-weight Pauli strings, and the site-occupation symmetry separates each
level into polynomially many small blocks.

The calculations illustrate both where the low levels of the hierarchy
succeed and where higher levels are needed. The full $(2,3)$-positivity
conditions are exact at the frustration-free Majumdar--Ghosh point and accurate
to within $0.08$ in units of the coupling, about one percent of the energy or
better, in the gapless and weakly frustrated regimes of
the Heisenberg ring and at the critical point of the transverse-field Ising
ring, while the spiral regime of the Heisenberg ring calls for higher levels.

Spin systems have been treated previously with variational 2-RDM methods,
including the Lipkin model~\cite{MazziottiErdahl01,GidofalviMazziotti06}, the
Hubbard model~\cite{HammondMazziotti06,Verstichel12}, the transverse-field Ising
model~\cite{Schwerdtfeger09}, and the XXZ model~\cite{RubioGarcia19}, as well as
with moment-based relaxations and the quantum
bootstrap~\cite{BaumgratzPlenio12,Haim20,Han20,Kull24,Gao25,Chadha26}. The
present work provides a complete hierarchy for qubits formulated directly as
constraints on the 2-RDM, that is, on the 2-qubit marginals. We use the
$(2,p)$-positivity conditions on the 2-RDM rather than the $p$-positivity
conditions of Mazziotti and Erdahl~\cite{MazziottiErdahl01} on the $p$-RDMs.
While the two are equivalent in how they constrain the 2-RDM, the
$(2,p)$-positivity conditions provide direct constraints on the 2-RDM. This
separation allows us to compute the 2-RDM without the higher RDMs and to exploit
structural features such as low rank in the polar-cone metric matrix $X$ of
Eq.~\eqref{eq:sdp-constraint}. Stationary-state
conditions such as variations of the contracted Schr\"odinger
equation~\cite{Mazziotti98,Mazziotti06ACSE} can be added to the semidefinite
program, as in Ref.~\cite{MazziottiErdahl01}, which may accelerate convergence
with the level of the hierarchy.

The same construction applies to other particle representations of qubits. In
two-mode representations, with one fermion~\cite{Abrikosov65} or one
boson~\cite{Schwinger65} in two modes per site, the qubit space is selected by
the vanishing site-occupation variances within a particle-conserving
framework, whereas one-mode representations, with hard-core
bosons~\cite{EmmanuelCosta26} or with fermions and Jordan--Wigner
strings~\cite{JordanWigner28}, require the particle-nonconserving
framework~\cite{Mazziotti26}. Each yields a hierarchy that converges to the same
set of qubit marginals, although the finite levels may differ.

Because the representability conditions hold for any quantum state, pure or
mixed, they also apply beyond ground states to the marginals measured on
quantum devices. Recent work has shown that imposing $N$-representability
conditions on fermionic RDMs measured on quantum devices, through constrained
shadow tomography~\cite{Avdic24,Avdic26} and correlated
purification~\cite{Wang26}, reduces the number of measurements and mitigates
statistical and hardware noise. The face correspondence extends these methods
from fermionic simulations to any quantum algorithm whose output is expressed
in terms of qubits: the 2-qubit marginals estimated from measurements can be
constrained by the $(2,p)$-positivity conditions whether or not the algorithm
simulates fermions, making qubit representability a general resource for error
mitigation, which we will explore in future work. More broadly, by identifying the
qubit marginal problem as a face of fermionic $N$-representability, this work
brings the constructive solution of the fermionic problem to quantum
information and simulation.

\begin{acknowledgments}
D.A.M. gratefully acknowledges the U.S. National Science Foundation Grant
No.~CHE-2155082.   The AI tool Claude Opus 5.5 High was used interactively for
scientific discussions and manuscript editing.  All scientific content was
independently evaluated and verified by the author, who takes full
responsibility for the results and conclusions.
\end{acknowledgments}

\noindent\textit{Data availability.}---The data that support the findings of this
article are openly available~\cite{DataRepo}.



\begin{thebibliography}{99}

\bibitem{Klyachko06}
A. A. Klyachko, Quantum marginal problem and $N$-representability,
J. Phys.: Conf. Ser. \textbf{36}, 72 (2006).

\bibitem{Schilling14}
C. Schilling, The quantum marginal problem, in \textit{Mathematical Results in
Quantum Mechanics} (World Scientific, Singapore, 2014).

\bibitem{Higuchi03}
A. Higuchi, A. Sudbery, and J. Szulc, One-qubit reduced states of a pure
many-qubit state: Polygon inequalities, Phys. Rev. Lett. \textbf{90}, 107902
(2003).

\bibitem{Liu06}
Y.-K. Liu, Consistency of local density matrices is QMA-complete,
arXiv:quant-ph/0604166.

\bibitem{LCV07}
Y.-K. Liu, M. Christandl, and F. Verstraete, Quantum computational complexity of
the $N$-representability problem: QMA complete, Phys. Rev. Lett. \textbf{98},
110503 (2007).

\bibitem{Yu21}
X.-D. Yu, T. Simnacher, N. Wyderka, H. C. Nguyen, and O. G\"uhne, A complete
hierarchy for the pure state marginal problem in quantum mechanics, Nat.
Commun. \textbf{12}, 1012 (2021).

\bibitem{Coleman63}
A. J. Coleman, Structure of fermion density matrices, Rev. Mod. Phys.
\textbf{35}, 668 (1963).

\bibitem{GarrodPercus64}
C. Garrod and J. K. Percus, Reduction of the $N$-particle variational problem,
J. Math. Phys. (N.Y.) \textbf{5}, 1756 (1964).

\bibitem{Kummer67}
H. Kummer, $N$-representability problem for reduced density matrices,
J. Math. Phys. (N.Y.) \textbf{8}, 2063 (1967).

\bibitem{Erdahl78}
R. M. Erdahl, Representability, Int. J. Quantum Chem. \textbf{13}, 697 (1978).

\bibitem{Mazziotti12}
D. A. Mazziotti, Structure of fermionic density matrices: Complete
$N$-representability conditions, Phys. Rev. Lett. \textbf{108}, 263002 (2012).

\bibitem{Mazziotti12PRA}
D. A. Mazziotti, Significant conditions for the two-electron reduced density
matrix from the constructive solution of $N$ representability, Phys. Rev. A
\textbf{85}, 062507 (2012).

\bibitem{Mazziotti23}
D. A. Mazziotti, Quantum many-body theory from a solution of the
$N$-representability problem, Phys. Rev. Lett. \textbf{130}, 153001 (2023).

\bibitem{Mazziotti26}
D. A. Mazziotti, Representability for quantum theory beyond particle-number
conservation, Phys. Rev. Lett. \textbf{136}, 120202 (2026).

\bibitem{MazziottiErdahl01}
D. A. Mazziotti and R. M. Erdahl, Uncertainty relations and reduced density
matrices: Mapping many-body quantum mechanics onto four particles, Phys. Rev. A
\textbf{63}, 042113 (2001).

\bibitem{Nakata01}
M. Nakata, H. Nakatsuji, M. Ehara, M. Fukuda, K. Nakata, and K. Fujisawa,
Variational calculations of fermion second-order reduced density matrices by
semidefinite programming algorithm, J. Chem. Phys. \textbf{114}, 8282 (2001).

\bibitem{Mazziotti02}
D. A. Mazziotti, Variational minimization of atomic and molecular ground-state
energies via the two-particle reduced density matrix, Phys. Rev. A \textbf{65},
062511 (2002).

\bibitem{Mazziotti04}
D. A. Mazziotti, Realization of quantum chemistry without wave functions
through first-order semidefinite programming, Phys. Rev. Lett. \textbf{93},
213001 (2004).

\bibitem{Zhao04}
Z. Zhao, B. J. Braams, M. Fukuda, M. L. Overton, and J. K. Percus, The reduced
density matrix method for electronic structure calculations and the role of
three-index representability conditions, J. Chem. Phys. \textbf{120}, 2095
(2004).

\bibitem{Cances06}
E. Canc\`es, G. Stoltz, and M. Lewin, The electronic ground-state energy
problem: A new reduced density matrix approach, J. Chem. Phys. \textbf{125},
064101 (2006).

\bibitem{Shenvi10}
N. Shenvi and A. F. Izmaylov, Active-space $N$-representability constraints for
variational two-particle reduced density matrix calculations, Phys. Rev. Lett.
\textbf{105}, 213003 (2010).

\bibitem{Poelmans15}
W. Poelmans, M. Van Raemdonck, B. Verstichel, S. De Baerdemacker, A. Torre,
L. Lain, G. E. Massaccesi, D. R. Alcoba, P. Bultinck, and D. Van Neck,
Variational optimization of the second-order density matrix corresponding to a
seniority-zero configuration interaction wave function, J. Chem. Theory Comput.
\textbf{11}, 4064 (2015).

\bibitem{Mazziotti16}
D. A. Mazziotti, Enhanced constraints for accurate lower bounds on
many-electron quantum energies from variational two-electron reduced density
matrix theory, Phys. Rev. Lett. \textbf{117}, 153001 (2016).

\bibitem{Alcoba18}
D. R. Alcoba, A. Torre, L. Lain, G. E. Massaccesi, O. B. O\~na, E. M. Honor\'e,
W. Poelmans, D. Van Neck, P. Bultinck, and S. De Baerdemacker, Direct
variational determination of the two-electron reduced density matrix for
doubly occupied-configuration-interaction wave functions: The influence of
three-index $N$-representability conditions, J. Chem. Phys. \textbf{148},
024105 (2018).

\bibitem{Mazziotti20}
D. A. Mazziotti, Dual-cone variational calculation of the two-electron reduced
density matrix, Phys. Rev. A \textbf{102}, 052819 (2020).

\bibitem{Knight22}
M. J. Knight, H. M. Quiney, and A. M. Martin, Reduced density matrix approach to
ultracold few-fermion systems in one dimension, New J. Phys. \textbf{24},
053004 (2022).

\bibitem{Knight25}
M. J. Knight, H. M. Quiney, and A. M. Martin, Reduced density matrix approach to
one-dimensional ultracold bosonic systems, arXiv:2503.15811.

\bibitem{Schouten25}
A. O. Schouten, S. Ewing, and D. A. Mazziotti, Two-electron reduced density
matrix method for the electronic structure of correlated materials, Phys. Rev.
B \textbf{112}, 235130 (2025).

\bibitem{Xie22}
J. Xie, S. Ewing, J.-N. Boyn, A. S. Filatov, B. Cheng, T. Ma, G. L. Grocke,
N. Zhao, R. Itani, X. Sun, H. Cho, Z. Chen, K. W. Chapman, S. N. Patel,
D. V. Talapin, J. Park, D. A. Mazziotti, and J. S. Anderson, Intrinsic
glassy-metallic transport in an amorphous coordination polymer, Nature
(London) \textbf{611}, 479 (2022).

\bibitem{Safaei18}
S. Safaei and D. A. Mazziotti, Quantum signature of exciton condensation, Phys.
Rev. B \textbf{98}, 045122 (2018).

\bibitem{Torres24}
L. I. P. Torres, A. O. Schouten, and D. A. Mazziotti, Molecular origins of
exciton condensation in van der Waals heterostructure bilayers, Chem. Sci.
\textbf{15}, 20371 (2024).

\bibitem{Kais07}
S. Kais, Entanglement, electron correlation, and density matrices, in
\textit{Reduced-Density-Matrix Mechanics: With Application to Many-Electron
Atoms and Molecules}, Adv. Chem. Phys. Vol.~134, edited by D. A. Mazziotti
(Wiley, New York, 2007), pp.~493--535.

\bibitem{Schouten26}
A. O. Schouten and D. A. Mazziotti, Entanglement complexity in many-body
systems from positivity scaling laws, Phys. Rev. A \textbf{113}, L030401 (2026).

\bibitem{Piris21}
M. Piris, Global natural orbital functional: Towards the complete description
of the electron correlation, Phys. Rev. Lett. \textbf{127}, 233001 (2021).

\bibitem{BenavidesRiveros20}
C. L. Benavides-Riveros, J. Wolff, M. A. L. Marques, and C. Schilling, Reduced
density matrix functional theory for bosons, Phys. Rev. Lett. \textbf{124},
180603 (2020).

\bibitem{Abrikosov65}
A. A. Abrikosov, Electron scattering on magnetic impurities in metals and
anomalous resistivity effects, Physics Physique Fizika \textbf{2}, 5 (1965).

\bibitem{Schwerdtfeger09}
C. A. Schwerdtfeger and D. A. Mazziotti, Convex-set description of quantum
phase transitions in the transverse Ising model using reduced-density-matrix
theory, J. Chem. Phys. \textbf{130}, 224102 (2009).

\bibitem{Chakraborty15}
R. Chakraborty and D. A. Mazziotti, Sufficient condition for the openness of a
many-electron quantum system from the violation of a generalized Pauli
exclusion principle, Phys. Rev. A \textbf{91}, 010101(R) (2015).

\bibitem{Smart19}
S. E. Smart, D. I. Schuster, and D. A. Mazziotti, Experimental data from a
quantum computer verifies the generalized Pauli exclusion principle, Commun.
Phys. \textbf{2}, 11 (2019).

\bibitem{Rockafellar70}
R. T. Rockafellar, \textit{Convex Analysis} (Princeton University Press,
Princeton, 1970).

\bibitem{Erdahl07}
R. M. Erdahl, The lower bound method for density matrices and semidefinite
programming, in \textit{Reduced-Density-Matrix Mechanics: With Application to
Many-Electron Atoms and Molecules}, Adv. Chem. Phys. Vol.~134, edited by
D. A. Mazziotti (Wiley, New York, 2007), pp.~61--91.

\bibitem{Slater50}
M. Slater, Lagrange multipliers revisited, Cowles Commission Discussion Paper
No.~403 (1950).

\bibitem{VandenbergheBoyd96}
L. Vandenberghe and S. Boyd, Semidefinite programming, SIAM Rev. \textbf{38},
49 (1996).

\bibitem{Braams07}
B. J. Braams, J. K. Percus, and Z. Zhao, The T1 and T2 representability
conditions, in \textit{Reduced-Density-Matrix Mechanics: With Application to
Many-Electron Atoms and Molecules}, Adv. Chem. Phys. Vol.~134, edited by
D. A. Mazziotti (Wiley, New York, 2007), pp.~93--101.

\bibitem{Mazziotti11}
D. A. Mazziotti, Large-scale semidefinite programming for many-electron quantum
mechanics, Phys. Rev. Lett. \textbf{106}, 083001 (2011).

\bibitem{MajumdarGhosh69}
C. K. Majumdar and D. K. Ghosh, On next-nearest-neighbor interaction in linear
chain. I, J. Math. Phys. \textbf{10}, 1388 (1969); On next-nearest-neighbor
interaction in linear chain. II, \textit{ibid.} \textbf{10}, 1399 (1969).

\bibitem{GidofalviMazziotti06}
G. Gidofalvi and D. A. Mazziotti, Computation of quantum phase transitions by
reduced-density-matrix mechanics, Phys. Rev. A \textbf{74}, 012501 (2006).

\bibitem{HammondMazziotti06}
J. R. Hammond and D. A. Mazziotti, Variational reduced-density-matrix
calculation of the one-dimensional Hubbard model, Phys. Rev. A \textbf{73},
062505 (2006).

\bibitem{Verstichel12}
B. Verstichel, H. van Aggelen, W. Poelmans, and D. Van Neck, Variational
two-particle density matrix calculation for the Hubbard model below half
filling using spin-adapted lifting conditions, Phys. Rev. Lett. \textbf{108},
213001 (2012).

\bibitem{RubioGarcia19}
A. Rubio-Garc\'{\i}a, J. Dukelsky, D. R. Alcoba, P. Capuzzi, O. B. O\~na,
E. R\'{\i}os, A. Torre, and L. Lain, Variational reduced density matrix method
in the doubly-occupied configuration interaction space using four-particle
$N$-representability conditions: Application to the XXZ model of quantum
magnetism, J. Chem. Phys. \textbf{151}, 154104 (2019).

\bibitem{BaumgratzPlenio12}
T. Baumgratz and M. B. Plenio, Lower bounds for ground states of condensed
matter systems, New J. Phys. \textbf{14}, 023027 (2012).

\bibitem{Haim20}
A. Haim, R. Kueng, and G. Refael, Variational-correlations approach to quantum
many-body problems, arXiv:2001.06510.

\bibitem{Han20}
X. Han, Quantum many-body bootstrap, arXiv:2006.06002.

\bibitem{Kull24}
I. Kull, N. Schuch, B. Dive, and M. Navascu\'es, Lower bounding ground-state
energies of local Hamiltonians through the renormalization group, Phys. Rev. X
\textbf{14}, 021008 (2024).

\bibitem{Gao25}
Q. Gao, R. A. Lanzetta, P. Ledwith, J. Wang, and E. Khalaf, Bootstrapping the
quantum Hall problem, Phys. Rev. X \textbf{15}, 031034 (2025).

\bibitem{Chadha26}
N. Chadha, M. G. Scheer, and E. Khalaf, Bootstrap bounds for quantum spin
systems using string operators, arXiv:2606.06584.

\bibitem{Mazziotti98}
D. A. Mazziotti, Contracted Schr\"odinger equation: Determining quantum
energies and two-particle density matrices without wave functions, Phys. Rev. A
\textbf{57}, 4219 (1998).

\bibitem{Mazziotti06ACSE}
D. A. Mazziotti, Anti-Hermitian contracted Schr\"odinger equation: Direct
determination of the two-electron reduced density matrices of many-electron
molecules, Phys. Rev. Lett. \textbf{97}, 143002 (2006).

\bibitem{Schwinger65}
J. Schwinger, On angular momentum, in \textit{Quantum Theory of Angular
Momentum}, edited by L. C. Biedenharn and H. van Dam (Academic Press, New York,
1965).

\bibitem{EmmanuelCosta26}
D. Emmanuel-Costa and M. Epping, Hard-core bosons in action: Applications to
quantum circuits, arXiv:2606.28004.

\bibitem{JordanWigner28}
P. Jordan and E. Wigner, \"Uber das Paulische \"Aquivalenzverbot, Z. Phys.
\textbf{47}, 631 (1928).

\bibitem{Avdic24}
I. Avdic and D. A. Mazziotti, Fewer measurements from shadow tomography with
$N$-representability conditions, Phys. Rev. Lett. \textbf{132}, 220802 (2024).

\bibitem{Avdic26}
I. Avdic, Y. Wang, M. Rose, L. I. Payne Torres, A. O. Schouten, K. J. Sung, and
D. A. Mazziotti, Constrained shadow tomography for molecular simulation on
quantum devices, Chem. Sci. \textbf{17}, 9633 (2026).

\bibitem{Wang26}
Y. Wang, I. Avdic, M. Rose, L. I. Payne Torres, A. O. Schouten, K. J. Sung, and
D. A. Mazziotti, Correlated purification for restoring $N$-representability in
quantum simulation, Phys. Rev. A \textbf{113}, 032424 (2026).

\bibitem{DataRepo}
Data for Solution of the qubit marginal problem from fermionic
representability, GitHub repository (2026),
\url{https://github.com/damazz/qubit-marginal-problem-from-fermionic-representability}.

\end{thebibliography}
\end{document}